\documentclass[11pt]{article}
\usepackage{moriond,epsfig}

\def\be{\begin{equation}}
\def\ee{\end{equation}}
\def\bea{\begin{eqnarray}}
\def\eea{\end{eqnarray}}

\begin{document}
\def\thefootnote{\fnsymbol{footnote}}
\setcounter{footnote}{1}
\vspace*{4cm}
\title{Energy loss and $x_2$ scaling breakdown in $J/\psi$ nuclear production~\footnote{Talk given at XXXVth Rencontres de Moriond: QCD and High Energy Hadronic Interactions, Les Arcs, France, 18-25 March 2000.}}

\author{F. Arleo~\footnote{email : arleo@subatech.in2p3.fr}, P.-B. Gossiaux, T. Gousset, and J. Aichelin}

\address{SUBATECH \\
Laboratoire de Physique Subatomique et des Technologies Associ\'ees\\
UMR Ecole des Mines de Nantes, IN2P3/CNRS, Universit\'e de Nantes\\
4, rue Alfred Kastler,
F-44070 Nantes Cedex 03, France.}

\maketitle\abstracts{
In addition to the final-state interactions, $J/\psi$ nuclear production might also be affected by parton energy loss. Using the upper limits from Drell-Yan data at SPS and Fermilab energies, we estimated energy loss contribution to $J/\psi$ production in $p$--$A$ collisions. The results indicated that the effects might be sizeable at 200~GeV while remaining small at higher energies.}

\setcounter{footnote}{0}
\section{Introduction}

Charmonium production is one of the cleanest hard probes studied in
high-energy hadron-hadron, hadron-nucleus, and nucleus-nucleus
scatterings. If we look at the reaction in the center of mass frame, the parton model tells us that the production results from the fusion
of a beam parton with energy-momentum fraction $x_1$ and a target
parton with energy-momentum fraction $x_2$ which are converted into a
$c\bar{c}$ pair that eventually gives the observed charmonium. In
the absence of nuclear effects the integrated production rate in
$A_1$--$A_2$ collision would be given by $A_1\times A_2$
times that in $p$--$p$ reaction. The deviation from this is often
interpreted within a conventionnal absorption model.

Recently the E866 collaboration provided the first evidence
for a distinction between $J/\psi$ and $\psi'$ in the differential
production ratio~\cite{lei00}
\begin{equation}\label{ratio}
R=\frac{{\mathrm d}\sigma(p+\mathrm{W}\to\psi+X)/\mathrm{d}x_F}
{A\,{\mathrm d}\sigma(p+\mathrm{Be}\to\psi+X)/{\mathrm d}x_F}.
\end{equation}
The distinction occurs in the kinematical region where the charmonia
have relatively small energies $E_{c\bar c}$ in the target rest
frame. This is a definite departure from the observation of identical ratios
for larger $c\bar c$ energies. It is tempting to describe this result by
saying that for small $c\bar c$ energies the nucleus ``acts as a
detector'' for the formation of the charmonium states. A possible
interpretation of this is that for large $c\bar c$ energies a
\emph{common} precursor to $J/\psi$ and $\psi'$ passes through the
target and final-state formation takes place outside the nuclear
volume. At low $c\bar c$ energies formation occurs within the target, leading to a weaker absorption for $J/\psi$ than $\psi'$ due to the smaller size of the former. At the time E866 data became available we decided to take these ideas
at face value and to test them in a simple but quantitative
scenario of nuclear absorption~\cite{arl00}.  

A direct consequence of such an approach is an $x_2$ scaling.  At high
energy the charmonium energy amounts to $E_{c\bar c}=x_1 E$ in the
target rest frame where $E$ is the beam energy in this frame. In terms
of the charmonium invariant mass $M_{c\bar{c}}$ one has $E_{c\bar c}=M_{c\bar c}^2/(2m_p x_2)$. Consequently, if the production ratio is in actual
facts driven by $E_{c\bar c}$ one expects $R$ to be independent of the
beam energy at a given $x_2$ ($x_2$ scaling).

Unlike what we expected, a comparison of $J/\psi$ differential production between 800~GeV (E866)~\cite{lei00} and earlier less precise 200~GeV data (NA3)~\cite{bad83} as a function of $x_2$ seems to indicate that 
$c\bar{c}$ production does not follow an $x_2$ scaling. However, it could happen that NA3 data reveal that at intermediate energies there is a mechanism involved in $c\bar{c}$ nuclear production beside the above mentionned nuclear suppression.

In fact, it has been argued for a long time that an incoming parton might lose some energy while scattering through nuclear matter. Its momentum fraction is then shifted from $x_1$+$\Delta x_1$ to $x_1$ at the point of fusion resulting in a modification of the effective beam parton density by a factor $F(x_1+\Delta x_1)/F(x_1)$. Because the parton distributions $F(x)$ drop dramatically at intermediate and large $x$, the effective density of beam partons and consequently the production ratio might decrease significantly.

The aim of these proceedings is to investigate whether energy loss could be a relevant mechanism to explain a breakdown of $x_2$ scaling in $J/\psi$ production. 

\section{Methods}

Since a corresponding quark energy loss may be expected in Drell-Yan (DY) pair production which is moreover not overwhelmed by final-state interactions, we first used the DY process to extract limits on parton energy loss. Energy loss is then combined with $c\bar{c}$ absorption to examine the consequences in the $J/\psi$ channel.

To be more specific we chose to test two alternative energy loss
scenarios. The first is that put forward by Gavin and
Milana~(GM)~\cite{gm}. It predicts a beam parton shift in momentum fraction
\begin{equation}\label {gm}
\Delta x_1=\kappa_1 x_1 A^{1/3},
\end{equation}
where $\kappa_1$ is a free parameter.\footnote{$\kappa_1$ might depend
on $Q^2$ and we needed its value in the region around $m_\psi^2$. It
turns out that the DY data are not precise enough and do not probe a
sufficient range in $Q^2$ to make any definite statement about the
latter dependence. We thus simply disregarded it.} We stress that the
shift depends on kinematics through $x_1$.

The second model was proposed by Brodsky and Hoyer~\cite{bh} in order to fix a potential defect in GM. On general grounds they showed that
\begin{equation}\label{bound}
\Delta x_1\le\frac{\kappa_2}{s} A^{1/3}, 
\end{equation}
with $\kappa_2\sim 0.5$~GeV$^2$, an upper bound with which $\Delta
x_1$ as given by Eq.~(\ref{gm}) is incompatible at large energies. For
accessible energies it also sets an upper bound on $\kappa_1$.
In relation with these considerations an alternative scenario, denoted BH in the following, is then obtained by assuming a shift in $x_1$ that saturates the bound given by~(\ref{bound}), i.e.,
$$
\Delta x_1=\frac{\kappa_2}{s} A^{1/3},
$$
where it should be noted here that the dependence on kinematics occurs
via the center of mass energy.

The free parameter in each model was fixed with DY data. The data set
encompassed, on the one hand, the 150 and 280~GeV pion beam data on $A=1$ and
195 targets~\cite{cal81} and, on the other hand, the 800~GeV proton beam data on
$A=9$ and 184~\cite{vas99}. At 800~GeV the data set spans a range in $x_2$ where
shadowing is known to be sizeable. We thus used data points corrected
for shadowing as given in Ref.~7. At SPS energies the data points lie in the
(weak) antishadowing region and the corresponding correction is
small. Calculations were done using MRST parton distributions~\cite{mrst} and shadowing corrections were given by the EKS parameterization~\cite{eks}.

The models were then used to assess the possible importance of energy
loss in $J/\psi$ production at 200 and 800~GeV. We assumed that $J/\psi$ production results from gluon fusion and took into account the color factor correction to energy loss~: $\Delta x_g = \frac{9}{4} \Delta x_q$~\footnote{At large $x_1$, $q\bar{q}$ annihilation into $c\bar{c}$ becomes predominant~\cite{vog00}. We checked that taking this channel into account did not change the overall results.}.
Subsequent energy loss was combined with nuclear absorption extracted from 800~GeV measurements and the result was compared to 200~GeV data~\cite{bad83}. In the $J/\psi$ sector shadowing corrections were not considered (see discussion at the end of Section~3).

\section{Results and Discussion}

\emph{Drell-Yan production}

In view to quantify parton energy loss, we first considered Drell-Yan
dimuon production. At 800~GeV the wide kinematic acceptance ($0.2\le
x_1\le 0.95$) and high statistics of E866/NuSea data make it possible to put stringent constraints on parton energy loss~\cite{vas99}. The fitted parameter values were
$$
\kappa_1=3\times 10^{-4}\quad(\chi^2/\mathrm{ndf}=0.69),
$$
for GM and
$$
\kappa_2=0.08\ \mathrm{GeV}^2\quad(\chi^2/\mathrm{ndf}=0.72),
$$
for BH. Their 1$\sigma$ upper limits, i.e., $\chi^2/$ndf$=1$, were 
$$
\kappa_1+\Delta\kappa_1=11\times 10^{-4},
$$
and
$$
\kappa_2+\Delta\kappa_2=0.77\ \mathrm{GeV}^2.
$$
This suggested that the Drell-Yan data --- corrected for shadowing --- are
consistent with zero energy loss within both models, but the 1$\sigma$
upper-limits imply that energy loss effects at a level of a few
hundreds of MeV$/$fm cannot be excluded.

Similarly, limits were estimated from a fit to the $\pi^-$ beam NA3
data at 150 and 280~GeV~\cite{cal81} (Table~\ref{NA3}). The lowest
$\chi^2/$ndf were reached with essentially no energy loss for both
models --- with or without inclusion of shadowing. Table~\ref{NA3}
also indicates that the consideration of shadowing effects tended to
improve the fits and correlatively to raise the upper limits of parton energy loss by a factor of four.

\begin{table}[htbp]
\begin{center}
\begin{tabular}{|c|c|c|c|c|}
\hline
 & \multicolumn{2}{|c|}{\centering GM loss}
 & \multicolumn{2}{|c|}{\centering BH loss}\\
\cline{2-5} 
 &{\centering $\kappa_1$ ($\times 10^{-3}$)} & $\chi^2/$ndf
 &\hspace{0.32cm}{\centering $\kappa_2$ (GeV$^2$)}\hspace{0.32cm}
 & \hspace{0.32cm}{\centering $\chi^2/$ndf}\hspace{0.32cm} \\
\hline
 no shadowing & 0.0 (1.0) & 0.86 & 0.0 (0.2) & 0.86\\
\hline
 with shadowing & 0.2 (3.7) & 0.43 & 0.0 (0.9) & 0.43\\
\hline
\end{tabular}
\end{center}
\caption{Energy loss parameters $\kappa_1$ and $\kappa_2$ from a fit
to NA3 data, with and without shadowing corrections. The 1$\sigma$
upper limits $\kappa_1+\Delta\kappa_1$ and $\kappa_2+\Delta\kappa_2$
are given between brackets.}
\label{NA3}
\end{table}

Fitting the nuclear dependence of Drell-Yan production in $p$ -- $A$
and $\pi$ -- $A$ reactions leads to the conclusion that quark
energy loss {\it has} to be small if any. Contrary to what we
expected, the comparison between 200 and 800~GeV data did not allow us
to discriminate between the different energy behaviours of GM and BH
scenarios. A closer look at the data showed that this is mostly due to the
large error bars in NA3 data.

\vspace{0.5cm}
\noindent{\emph{$J/\psi$ production}}

In the parton model, production of $c\bar{c}$ pairs proceeds via parton fusion and might therefore be affected by initial-state effects (e.g., energy loss and
shadowing), as well as final-state interactions. From the preliminary
Drell-Yan study, we assumed $\kappa_1$ to be equal to 0.001 and
$\kappa_2$ to be equal to 0.5 GeV$^2$ to evaluate the possible energy loss
contribution to $J/\psi$ production.

\begin{figure}[htbp]
\centerline{\psfig{figure=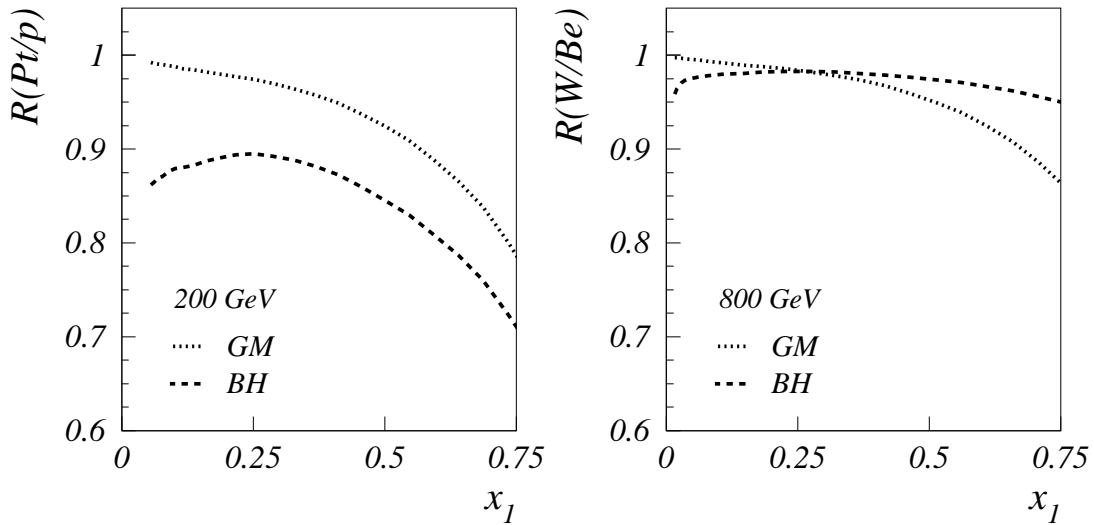,width=14.8cm}}
\caption{Energy loss effects on $J/\psi$ production according to GM
loss ({\it dotted}) and BH loss ({\it dashed}) at 200~GeV ({\it left})
and 800~GeV ({\it right}). The production ratio is defined as $R(A/B)
\equiv B\sigma(A)/A\sigma(B)$.}
\label{wlonly}
\end{figure}

Figure~\ref{wlonly} shows the $J/\psi$ production ratios $R(\mathrm{Pt}/p)$ at 200~GeV and $R(\mathrm{W}/\mathrm{Be})$ at 800~GeV. The most striking result is the
large suppression seen at SPS according to BH mechanism whereas this suppression remains negligible at Fermilab energies. The
$J/\psi$ production at 200~GeV is already affected ($R\approx$~0.9) at
low $x_1$, and decreases when $x_1$ increases. The situation is
different in the GM approach where the loss scales with $x_1$,
independently of the beam energy. This can be seen in
Figure~\ref{wlonly}~\footnote{The difference between the two dotted
curves in Figure~\ref{wlonly} arises from the normalization with $p$ and
Be targets at 200~GeV and 800~GeV respectively.} which also
demonstrates that the effect is moderate for $\kappa_1=0.001$ with a
decrease from 1 at small $x_1$ to $0.8$ at $x_1=0.7$. The decrease
seen in both approaches when going from small to large $x_1$ is the
behaviour that might be reflected by NA3 $J/\psi$ data.

\begin{figure}[htbp]
\centerline{\psfig{figure=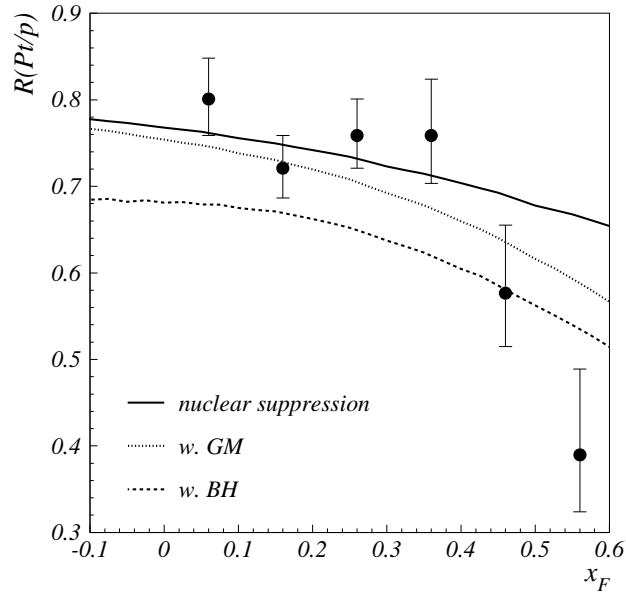,width=8.5cm}}
\caption{Ratio of $J/\psi$'s produced in $p$(200 GeV)+Pt and
$p$(200~GeV)+$p$ versus $x_F$ from NA3 as compared to theoretical calculations of the nuclear
suppression ({\it solid}), with additional GM energy loss ({\it dotted}) or additional BH energy loss
({\it dashed}).}
\label{figna3}
\end{figure}

Once energy loss in the $J/\psi$ sector was examined, $J/\psi$
suppression was calculated taking in addition nuclear absorption into
account. The results are compared with experimental data~\cite{bad83} on
Figure~\ref{figna3} as a function of $x_F \equiv x_1 - x_2$. First, Figure~\ref{figna3}
reminds us that the model described in Ref.~2 for charmonium suppression (solid curve) is unable to
describe large $x_F$ NA3 data. Second, the energy loss mechanism
proved to reduce the disagreement with large $x_F$ data points. In
particular, GM loss did not affect low $x_F$ production ratios and somehow
decreased $J/\psi$ production at $x_F\approx 0.5$. By
contrast, the BH scenario could almost reproduce the
large suppression seen at large $x_F$ but tended to give too strong a
suppression at small $x_F$. Though it is true that a first level
comparison of both models would favor the GM approach,
we should however stress that the results depend strongly on the
values chosen for the parameters $\kappa_1$ and $\kappa_2$. This
prompted us not to draw too quantitative conclusions.

In summary, the results obtained may indicate that NA3 $J/\psi$ data
result from a possible interplay between energy loss and nuclear
suppression. In particular, the general trend of data is better
reproduced when parton energy loss is taken into account. Consequently, the
lack of $x_2$ scaling exhibited by NA3 and E866/Nusea data could be
---~at least partly~--- explained with such a mechanism. One
limitation nevertheless arises from the remaining discrepancy between data 
and theoretical calculations. Further to that, we
should mention that other mechanisms have been neglected in this
study. In particular, gluon antishadowing calculated with EKS98 leads
to an enhancement of $J/\psi$ production in platinum near $x_2
\approx$~0.1 (i.e.  $x_F \approx$~0.2 for NA3).

It is hardly necessary to repeat that precise measurements of both
Drell-Yan and charm differential production over a large kinematic
window is mandatory to disentangle all the mechanisms
involved in $J/\psi$ production.

{}

\end{document}